\documentclass[final]{emulateapj}
\usepackage{apjfonts}
\usepackage{amssymb,latexsym,amsmath}
\shorttitle{Where Are All The Fallback Disks?}
\shortauthors{Ek\c{s}{i} et al.}

\begin{document}

\title{Where Are All The Fallback Disks? Constraints on Propeller Systems}

\author{Ek\c{s}{i}, K.Y., Hernquist, L. \and Narayan, R.}
\affil{Harvard-Smithsonian Center for Astrophysics, \\
60 Garden Street, Cambridge, MA 02138}


\email{yeksi@cfa.harvard.edu, lhernqui@cfa.harvard.edu, narayan@cfa.harvard.edu}

\begin{abstract}

Fallback disks are expected to form around new-born neutron stars
following a supernova explosion.  In almost all cases, the
disk will pass through a propeller stage. If the neutron star is
spinning rapidly (initial period $\sim 10$ ms) and has an ordinary
magnetic moment ($\sim 10^{30}$ G cm$^3$), the rotational power
transferred to the disk by the magnetic field of the neutron star will
exceed the Eddington limit by many orders of magnitude, and the disk
will be rapidly disrupted.  Fallback disks can thus survive only
around slow-born neutron stars and around black holes, assuming the
latter do not torque their surrounding disks as strongly as do neutron
stars.  This might explain the apparent rarity of fallback disks
around young compact objects.

\end{abstract}


\section{Introduction}

Following a supernova explosion, some of the ejected matter may remain
bound to the remnant and fall back \citep{colgate}.  A disk can form
from this material if its specific angular momentum $l$ exceeds the
Keplerian value at the surface of a newly formed neutron star,
$l_{\text{K}}=\sqrt{GMR}\simeq 1.4\times 10^{16}$ cm$^{2}$ s$^{-1}$,
where $M$ is the mass of the star and $R$ is its radius.  Numerical
simulations of the pre-supernova evolution of rotating stars
\citep{heger} suggest $l\approx 10^{16}-10^{17}$ cm$^{2}$ s$^{-1}$,
implying that disk formation is possible 
\citep[see also ][]{minesh97}.

\citet{MD81} suggested that fallback disks may be an ingredient of
radio pulsars, as an alternative model to the conventional view that
pulsars are isolated electromagnetic engines.  \citet{LWB} associated
the formation of planets around pulsars with fallback
disks. \citet{meyer89} and \citet{meyer92} invoked a fallback disk
around the remnant of \objectname{SN 1987A} to explain the deviation
of the observed light curve from the theoretical light curve for pure
radioactive decay \citep[see also ][]{mineshige93}.

More recently, fallback disks \citep{CHN,alpar01,marsden01a,EA03} have
been invoked to explain anomalous X-ray pulsars \citep[AXPs; ][]{WT04}
via an accretion model rather than the competing magnetar model
\citep{DT92}.  This model has rekindled interest in the possible
implications of fallback disks around radio pulsars.  A fallback disk
assisting magnetic dipole radiation torque can explain the age
discrepancy \citep{marsden01b} and braking indices \citep{menou01}
of pulsars, as well as the distribution of radio pulsars on the
$P-\dot{P}$ diagram \citep{AAY01}. \citet{BP04} suggested that the
jets observed in the \objectname{Crab} and \objectname{Vela} pulsars
are collimated by fallback disks. \citet{qiao03} proposed a fallback
disk model for periodic timing variations of pulsars. If there are
fallback disks assisting the magnetic dipole torque, field estimates
based on pure dipole radiation ($B\approx 6\times
10^{19}(P\dot{P})^{1/2}$) would be overestimates.  Note that the
so-called high magnetic field pulsars
\citep{gonza03,gonza04,mclaugh03a,mclaugh03b} have braking indices less than
three, indicating that they are not spinning down purely by emitting
magnetic dipole radiation.

A fallback disk \citep{xuwang,shi03} has also been invoked for the
central compact objects \citep[CCOs; ][]{pavlov04} in supernova
remnants.  These objects are likely relatives of AXPs. A typical
member of this small group is the central X-ray point source in
Cassiopeia A which is just 320 years old and hence likely to possess a
fallback disk. A fallback disk model of soft gamma ray repeaters
\citep[SGRs; ][]{WT04} was invoked by \citet{ertan03} who showed that
the X-ray enhancement following the August 27 giant flare of
\objectname{SGR 1900+14} can be modeled as the post-burst relaxation
of the inner region of a disk which had been pushed back by the flare
energy.

While there is ample theoretical motivation for considering fallback
disks around young compact objects, the existence of these disks has
never been confirmed observationally.  There are tight limits on the
luminosities of young systems like SN87A \citep{park04} and Cas A
\citep{chakra} where fallback disks are especially likely. For the
latter object, \citet{chakra} showed that the optical properties of
the source are not similar to those of quiescent low-mass X-ray
binaries.  Infrared emission is predicted \citep{PHN} by accretion
models of AXPs, and possible infrared counterparts have even been
identified \citep[see ][ and references therein]{tam04} for five of
the six known AXPs.  The observations are, however, not consistent
with a standard thin disk model.  Moreover, observations of bursts
from AXPs \citep{gav02} suggest that AXPs are relatives of SGRs for
which the magnetar model is favored.

In this \emph{letter} we address the lack of observational evidence
for fallback disks. We argue that fallback disks will be short lived
because they would be disrupted by the spin-down power of the neutron
star, either in the propeller phase
\citep{shv70a,IS75,DP81,romanova04} or in the radio pulsar phase.  In
the following section we calculate the spin-down power
\citep{pri86,sha86,tre87} transferred from the magnetosphere of a
neutron star to the disk in the propeller regime and show that for
conventional initial periods $P_{\text{0}} \sim 10$ ms and magnetic
moments $\mu\sim10^{30}$ G cm$^3$, as estimated for radio pulsars, the
luminosity of the disk will be much larger than the Eddington limit
for all mass flow rates. In this regime, accretion is highly unstable,
and the disk is likely to evolve very quickly until the inner radius
of the disk is outside the light cylinder (see \S~2).  Beyond this
point, the neutron star becomes a radio pulsar and the star can no
longer torque the disk (or vice versa).

In the context of wind-fed, mass-exchange binaries, \citet{shv70b}
argued that a fast-rotating neutron star would first become an ejector
(radio pulsar), in which the disk remains outside the light cylinder.
Only after the star slows down can inflowing matter penetrate the
light cylinder, allowing the propeller stage to commence.  In the case
of fallback disks, a fixed amount of mass is available in the disk,
rather than a continuous supply of matter as in a mass-exchange
binary.  Therefore, once a rapidly rotating neutron star progresses
through the initial unstable accretion and propeller stages to an
ejector/radio pulsar stage, it is the end of the road for the disk;
the system cannot switch back to a propeller or accretion phase.  We
elaborate upon these points in the following sections.

\section{Constraints on a Propellers}

In order that a fallback disk affect the spin evolution of a neutron
star, the inner radius of the disk must be inside the light-cylinder
radius $R_{\text{L}}=c/\Omega$ where $\Omega$ is the stellar angular
velocity.  To within a factor of order unity, the inner radius of the
disk, $R_{\text{m}}$, is approximated by the Alfv\'en radius
\citep{DO73}:
\begin{equation}
R_{\text{m}}\cong R_{\text{A}}\equiv
\left(\frac{\mu^2}{\sqrt{2GM}\dot{M}} \right)^{2/7},
\label{inner}
\end{equation}
where $\dot{M}$ is the mass flow rate and $\mu$ is the stellar
magnetic moment.  When the disk is inside the light cylinder it
suppresses radio pulsar emission.  If the inner radius of the disk is
also inside the co-rotation radius
$R_{\text{c}}=(GM/\Omega^{2})^{1/3}$, the inflowing matter will
accrete and spin the star up.

Initially, a fallback disk is likely to form in the accretion phase.
As the $\dot{M}$ in the disk declines (recall that fallback disks are
not replenished as are disks in binary systems), the inner radius
moves out, and when it goes beyond the co-rotation radius, the system
will enter the propeller stage where the inflowing matter, instead of
being accreted, is expelled.  Here, the neutron star spins down
through the interaction of its magnetosphere with the disk.  If the
torque (angular momentum flux) acting on the star by the disk is $N$,
the power transferred to the disk by the neutron star will in turn be
$-\Omega N$.  This power will add to the gravitational power of the
inflowing matter, viz., $GM\dot{M}/R_{\text{m}}$, enhancing the energy
budget of the disk.  Some of the energy will be used to drive matter
from the system, provided the velocity of the expelled matter,
$v_{\text{out}}$, is greater than the escape velocity
$v_{\text{esc}}=\sqrt{2GM/R_{\text{m}}}$.  The outflowing matter will
carry away kinetic energy at the rate $(1/2)\dot{M}v_{\text{out}}^2$.
Hence, the net radiative luminosity of the disk in the propeller phase
is given by
\begin{equation}
L_{\text{disk}} =
\begin{cases}
GM\dot{M}/R_{\text{m}}-\Omega N-\dot{M}v_{\text{out}}^{2}/2 & \text{if $v_{\text{out}}>v_{\text{esc}}$,} \\
GM\dot{M}/R_{\text{m}}-\Omega N & \text{if $v_{\text{out}}<v_{\text{esc}}$.}
\end{cases}
\label{L_tot1}
\end{equation}

The velocity that the expelled matter can attain depends on the
coupling of the disk with the magnetosphere; i.e the torque, which can
be written as $N=n R_{\text{m}}^{2}\Omega
_{\text{K}}(R_{\text{m}})\dot{M}$. Here, $n$ is the dimensionless
torque and depends only on the dimensionless fastness parameter
$\omega_{\ast}=\Omega/\Omega_{\text{K}}(R_{\text{m}})$.  Conservation
of angular momentum requires that
\begin{equation}
v_{\text{out}}=R_{\text{m}}\Omega _{\text{K}}(R_{\text{m}})
(1- n).
\label{vout}
\end{equation}
Using this in equation~(\ref{L_tot1}) we find that the luminosity of the disk
in the propeller regime is
\begin{equation}
L_{\text{disk}}=\frac{GM\dot{M}}{R_{\text{m}}}
\begin{cases}
1-\omega_{\ast } n - (1-n)^{2}/2 & \text{if $v_{\text{out}}>v_{\text{esc}}$,} \\
1-\omega_{\ast } n  & \text{if $v_{\text{out}}<v_{\text{esc}}$.}
\end{cases}
\label{L_tot2}
\end{equation}
For $v_{\text{out}}<v_{\text{esc}}$ the matter expelled by the
magnetosphere can return to the disk at some radius larger than the
inner radius.  As shown by \citet{ST93}, the matter will then
accumulate near the inner boundary and accrete sporadically.

Many prescriptions have been discussed in the literature for propeller
torques. We estimate $n$ by assuming that the magnetosphere and the
inflowing matter are ``particles'' colliding and transferring angular
momentum.  If the collisions are elastic and the moment of inertia of
the star is much larger than that of the inflowing matter one finds
$n=2(1-\omega_{\ast})$, whereas completely inelastic collisions would
give $n=1-\omega_{\ast}$.  In both cases we assume that the mass of
the particle representing the magnetosphere is much greater than that
of the particle representing the accreting fluid.  Let us thus define
an ``elasticity parameter'' $\beta$ and write
\begin{equation}
n(\omega_{\ast}) =(1+\beta)(1-\omega_{\ast}),
\label{jdot}
\end{equation}
where $\beta$ varies between zero and unity and measures how
efficiently the kinetic energy of the neutron star is converted into
kinetic energy of expelled matter through the interaction of the
magnetosphere with the disk.  The limiting case of $\beta=0$
corresponds to a ``completely inelastic" interaction in which the
fraction converted to heat is the maximum, while $\beta=1$ corresponds
to an ``elastic'' interaction in which the rotational energy of the
neutron star is converted completely to kinetic energy of expelled
matter without heating the disk. This can be seen more clearly when we
substitute the torque prescription (\ref{jdot}) into
equation~(\ref{L_tot2}):
\begin{equation}
L_{\text{disk}}=\frac{GM\dot{M}}{2R_{\text{m}}}
\begin{cases}
1+( 1-\beta^2)\left(\omega_{\ast} -1\right) ^{2} & \text{if $v_{\text{out}}>v_{\text{esc}}$,} \\
2-2(1+\beta)(1-\omega_{\ast})\omega_{\ast }  & \text{if $v_{\text{out}}<v_{\text{esc}}$.}
\end{cases}
\label{L_tot3}
\end{equation}
For $\beta=1$, $L_{\text{disk}}=GM\dot{M}/2R_{\text{m}}$ which is
precisely the luminosity of a Keplerian disk with no torque applied at
the inner boundary.  Note that \citet{CHN} employed
$n=2(1-\omega_{\ast})$, referring to the detailed numerical
simulations of \citet{daumerie}.  This corresponds to the limiting
case of an ``elastic interaction'' ($\beta=1$) in which no heat is
produced in the disk by the energy transferred from the neutron star.
As any fluid process that strips matter from the disk would be
dissipative, the $\beta=1$ limit will never be realized in practice.
In the accretion regime, the tangential velocity of the flow will
adjust to the velocity of the magnetosphere, which corresponds to
$\beta=0$.  For the torque to be continuous across the transition from
the accretion to propeller regimes, we need to assume $\beta=0$ at
least for $\omega_{\ast}\approx 1$.  As the dependence of the
expressions in equation (\ref{L_tot3}) on $\beta$ is weak (so long as
$\beta$ is not too close to unity), choosing $\beta$ in the range
$0-0.5$ produces very little difference in the results.  Note that
early propeller torques employed in the literature were quite
inefficient. In the original work of \citet{IS75} the dimensionless
torque was estimated to be $1/\omega_{\ast}$.  In contrast, the recent
detailed simulations of \citet{romanova04} suggest even stronger
torques than we employ.

From equation~(\ref{L_tot3}), we can solve for the minimum period
$P_{\rm min_1}$ below which the luminosity of a propeller disk will
exceed the Eddington limit, $L_E=4\pi GMc/\kappa_{\text{es}}$, where
we take $\kappa_{\text{es}}=0.2$ for the electron scattering opacity
as appropriate for the high metal content of the fallback matter.
Taking $\beta=0$, and defining $\theta = L_{\text{E}}
R_{\text{m}}/GM\dot{M}$, we obtain
\begin{equation}
P_{\text{min$_1$}}=\frac{2\pi R_{\text{m}}^{3/2}}{\sqrt{GM}}
\begin{cases}
 \left(\sqrt{2\theta-1}+1 \right)^{-1} & \text{if $v_{\text{out}}>v_{\text{esc}}$,} \\
 \left(1/2+\sqrt{\theta-3/4} \right)^{-1}  & \text{if $v_{\text{out}}<v_{\text{esc}}$.}
\end{cases}
\label{pmin1}
\end{equation}
Another limiting period $P_{\rm min_2}$ is obtained by the condition
that the inner radius of the disk be inside the light cylinder.  This
gives
\begin{equation}
P_{\text{min$_2$}}=\frac{2\pi}{c}R_{\text{m}} \, .
\label{pmin2}
\end{equation}
Finally, if the disk is in the accretion phase, then the maximum $\dot
M$ is given by the Eddington limit: $GM\dot M_{\rm max}/R = L_{\rm
Edd}$.  Figure~\ref{fig_1} shows all these limits in the $\dot M - P$
plane.  We see that only a limited area of the plane is consistent
with the various constraints we have discussed.  Note that, for a
given stellar magnetic moment, there is a global minimum period
$P_{\rm min}$ below which no stable disk is permitted for \emph{any}
$\dot{M}$.  A neutron star that is born with a shorter period cannot
possibly have a stable fallback disk.  For $\mu=10^{30} ~{\rm
G\,cm^3}$, the global minimum period is $P_{\rm min}=35.2$ ms. For a
general $\mu$ and $R_{\text{L}} \gg R_{\text{c}}$, the minimum period can be
approximated as
\begin{equation}
P_{\rm min}\approx 0.036 \mu_{30}^{4/7} ~{\rm  s},
\label{pmin}
\end{equation}
where $\mu_{30}=\mu/10^{30}$.
\placefigure{fig_1} 

The propeller regime with super-critical mass flow rates has been
studied by \citet{min91} who argued that the inner radius of the disk
will adjust such that the luminosity does not exceed the Eddington
limit.  Our work shows that for a neutron star with standard
parameters, $\mu=10^{30}$ G cm$^{3}$, $P=10$ ms, there is no way for
the disk to penetrate the light cylinder and yet have its luminosity
be less than the Eddington limit.  In such a system, we argue that the
accretion and propeller phases will progress very rapidly, on a much
shorter time scale than the viscous time of the disk, as the accretion
and spin-down power cause the inner regions of the disk to evaporate.
Once the disk moves outside the light cylinder, conventional wisdom
\citep{shv70b} suggests that the disk would be disrupted by the
radiative pressure of the neutron star which now acts like a radio
pulsar.  Recently, \citet{EA05} argued that the transition from the
inner zone dipole field to the radiative zone field can be broad,
especially if the inclination angle between the rotation and magnetic
field axes is small, which would imply that the disk may be able to
find a stable inner boundary outside the light cylinder.  If this is
the case, the pulsar would evaporate the disk slowly by the
interaction of the pulsar wind with the disk.  Whether the pulsar will
be slowed down to periods where the inner radius of the disk can
re-penetrate the light cylinder depends on the complex physics of this
interaction.

\section{Discussion}

We have shown in this paper that fallback disks around new-born
neutron stars with conventional periods and magnetic moments would be
disrupted by a combination of the accretion power and the rotational
power of the neutron star being transferred to the disk.  Therefore,
even though fallback disks may well form, they would not be able to
survive for standard initial neutron star parameters.

\begin{figure}[t]
\epsscale{1.3} \plottwo{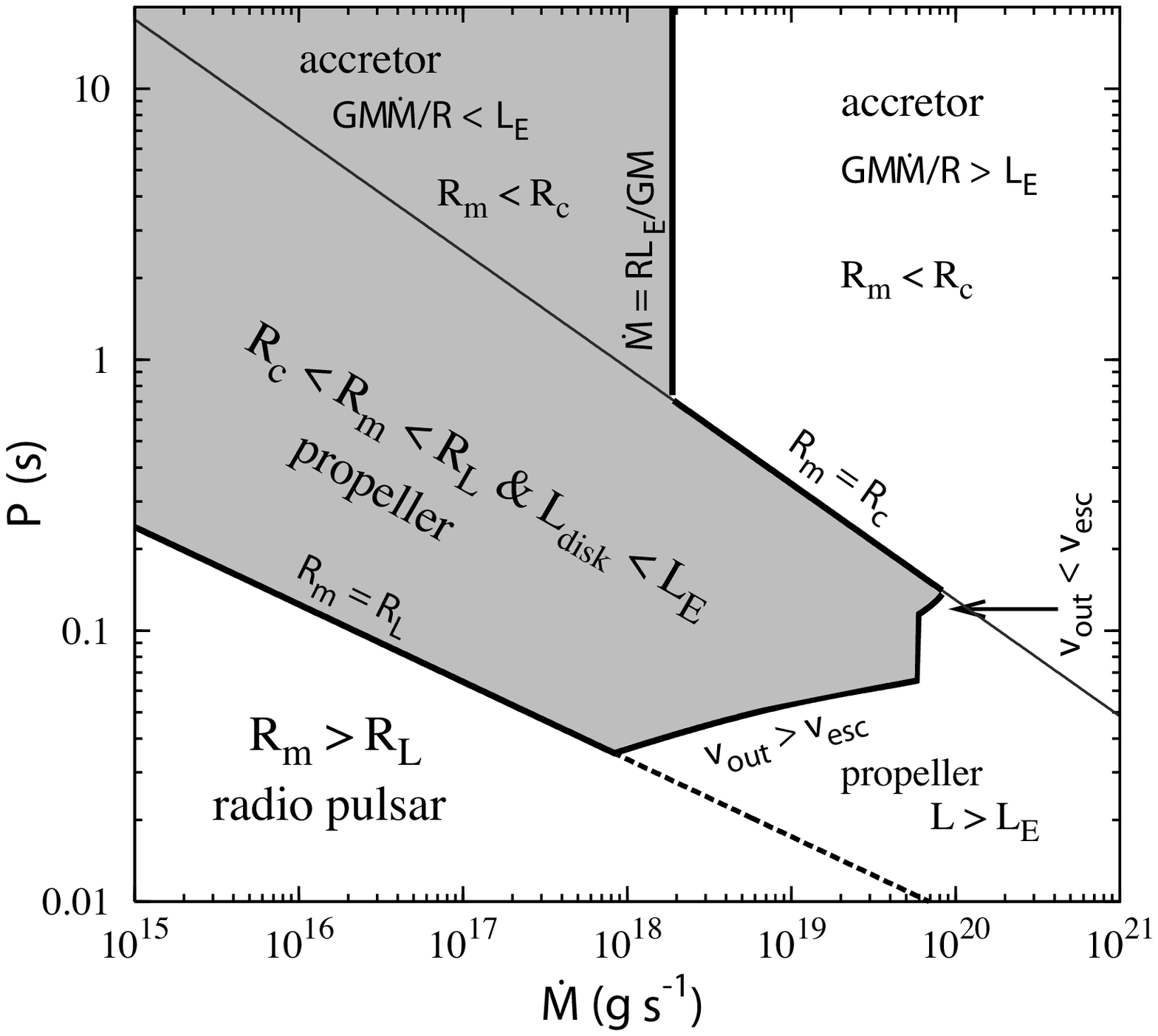}{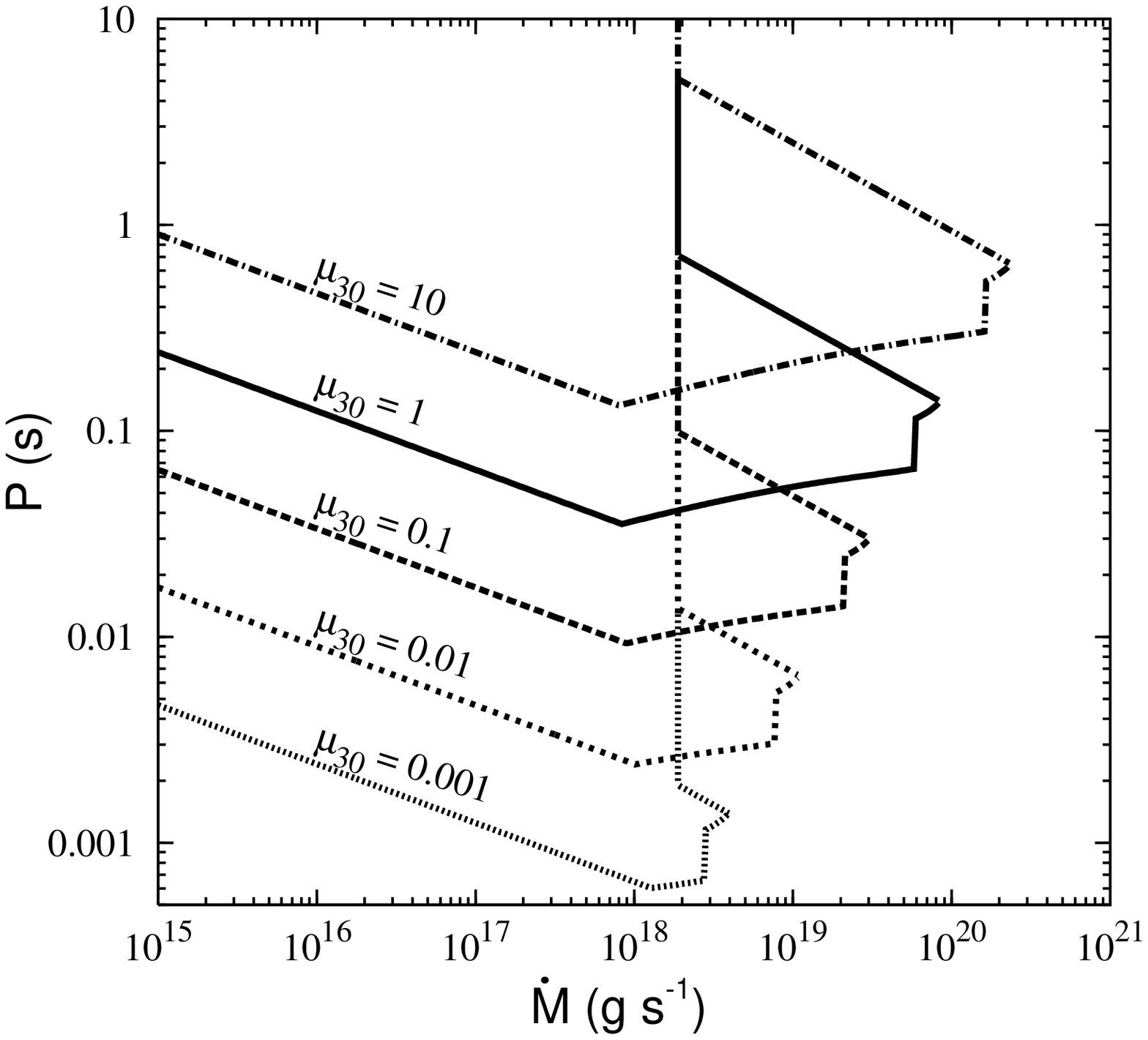}
\caption{First Panel: Allowed regions (shaded) and disallowed regions
(not shaded) in the (spin period $P$)--(mass accretion rate $\dot M$)
plane for an accretion disk around a spinning magnetized star with a
magnetic moment $\mu=10^{30} ~{\rm G\,cm^3}$.  The boundary between
the two regions is demarcated by various lines corresponding to (i)
$R_m=R_L$ which separates the radio pulsar and propeller phases, (ii)
$R_m=R_c$ which separates the propeller and accretion phases, and
(iii) $L=L_E$ which separates the sub-Eddington and super-Eddington
accretion phases.  Note that there is a minimum allowed period $P_{\rm
min}$ which is independent of $\dot M$, and a maximum accretion rate
which is independent of $P$.  Second Panel: Shows the boundaries
between the allowed and disallowed regions in the $P$--$\dot M$ plane
for several choices of $\mu$.  The minimum allowed period $P_{\rm
min}$ scales with $\mu$ as in equation (\ref{pmin}).}

\label{fig_1}
\end{figure}

Our results are general, in the sense that for all propeller systems
with parameters outside the shaded region in Figure~\ref{fig_1} the
disk is either outside the light cylinder or will be pushed there
rapidly.  Thus, not all combinations of the magnetic moment and period
are allowed.  In particular, for a given magnetic moment, there is a
minimum allowed neutron star period $P_{\rm min}$ given by equation
(\ref{pmin}).  If a neutron star is born with a shorter period that
this, only a very short-lived fallback disk is possible.  This might
explain why there has been no direct observational evidence for
fallback disks around young neutron stars even though one expects a
fair amount of fallback material with angular momentum to accumulate
during and soon after the supernova explosion.

The recent interest in the fallback disk model is because of the
potential application of this model to AXPs.  An advantage of this
model is that the neutron stars in AXPs are drawn from the same
population as radio pulsars, except for one new parameter describing
the initial mass of the disk \citep{chat}.  However, our present results indicate
that, if fallback disks are to be present around AXPs, these neutron
stars must have been born with long periods and/or small magnetic
dipole fields that are not typical of radio pulsars. It is interesting
that recent studies \citep{vra04} based on the Parkes multibeam survey
imply that 40\% of pulsars are ``injected'' with initial periods in
the range $0.1-0.5$ s \citep[as first suggested by ][ using a smaller
sample of pulsars] {vive81}.  For the fallback model to be relevant,
AXPs must be drawn from this sub-population of slow injected pulsars.

For SN87A, \citet{OA04} assumed that the compact star was born
spinning rapidly and that its present luminosity is due to pure
magnetic dipole spin-down.  They concluded that the observed upper
limit on the luminosity of the star requires the magnetic field of the
putative pulsar to be either very small (almost in the millisecond
pulsar range) or very large (in the magnetar range).  In the latter
case, the pulsar has spun down rapidly after the supernova explosion,
and so the present luminosity is low.  The magnetar solution branch of
\citet{OA04} falls in the region of Figure 1 in which a fallback disk
is not allowed.  Therefore, their solution is viable; even if the
system had fallback material it would have been ejected early on.
However, the second solution branch of \citet{OA04} corresponds to a
small magnetic dipole moment ($\mu_{30} \sim 0.01$) for the neutron
star, so that a fallback disk is very likely to survive.  The
luminosity from the disk in the propeller phase would then be much
larger than the magnetic dipole luminosity assumed by \citet{OA04}.
Therefore, the upper limit on the magnetic field must be much lower in
order to satisfy the observational constraint on the luminosity.

An alternative explanation for the lack of fallback disks is given by
\citet{FCP99}, who argued that accretion flows with heavy elements
would be dynamically unstable when the recombination of heavy elements
commences, giving rise to a line-driven wind that eventually expels
all the initially bound material.

Finally, we note that supernova explosions may sometimes lead to black
holes. The formation of fallback disks around black holes was
investigated by \citet{minesh97}, and a fallback disk model for
ultraluminous X-ray sources \citep[ULXs; ][]{FW04} was suggested by
\citet{li03}.  Rotating black holes may transfer energy to their
surrounding disks \citep[e.g., ][]{li02}, just like neutron stars, in
which case such disks would again tend to reach super-Eddington
luminosities and be disrupted.  However, if the energy transfer is
significantly less than in neutron stars, the disk might find it
easier to survive.  It would be interesting to search for fallback
disks around young black hole systems. \citet{li03} mentions two ULXs
associated with supernova remnants that could be such accreting black
holes.

\acknowledgements
We thank M.Ali Alpar and \"{U}nal Ertan for useful discussions.
This work was supported in part by NASA grant NNG04GL38G.

\newpage



\begin{thebibliography}{99}

\bibitem[Alpar(2001)]{alpar01} Alpar, M. A. 2001, \apj, 554, 1245.

\bibitem[Alpar et al.(2001)]{AAY01} Alpar, M. A., Ankay, A., \& Yazgan, E.
2001, \apj, 557, L61

\bibitem[Blackman \& Perna(2004)]{BP04} Blackman, E. G., \& Perna, R., 2004,
\apj, 601, L71

\bibitem[Chakrabarty et al.(2001)]{chakra} Chakrabarty, D., Pivovaroff, M. J., Hernquist, L. E., Heyl, J. S., \& Narayan, R., 2001, \apj, 548, 800

\bibitem[Chatterjee \& Hernquist(2000)]{chat} Chatterjee, P. \& Hernquist, L., \apj, 543, 368

\bibitem[Chatterjee et al.(2000)]{CHN} Chatterjee, P., Hernquist, L., \& Narayan, R. 2000, \apj, 534, 373

\bibitem[Colgate(1971)]{colgate} Colgate, S. A., 1971, \apj, 163, 221.

\bibitem[Daumerie(1996)]{daumerie} Daumerie, P., PhD thesis, 1996, Illinois Univ.

\bibitem[Davidson \& Ostriker(1973)]{DO73} Davidson, K., \& Ostriker, J. P., 1973, \apj, 179, 585

\bibitem[Davies \& Pringle(1981)]{DP81} Davies, R. E., \& Pringle, J. E., 1981,
\mnras, 196, 209.


\bibitem[Duncan \& Thompson(1992)]{DT92} Duncan, R.C., \& Thompson, C., 1992,
\apj, 392, L9

\bibitem[Ek\c si \& Alpar(2003)]{EA03} Ek\c si, K. Y., \& Alpar, M. A., 2003,
\apj, 599, 450.

\bibitem[Ek\c si \& Alpar(2005)]{EA05} Ek\c si, K. Y., \& Alpar, M. A., 2005, \apj, in press, astro-ph/0409408

\bibitem[Ertan \& Alpar(2003)]{ertan03} Ertan, Ü., \& Alpar, M. A., 2003, \apj, 593, L93


\bibitem[Fryer, Colgate \& Pinto(1999)]{FCP99} Fryer  C.L., Colgate S. A., \& Pinto P. A., 1999, \apj 511, 885

\bibitem[Fabbiano \& White(2004)]{FW04} Fabbiano, G. \& White, N.E., 2004, in
"Compact Stellar X-ray Sources", eds. W.H.G. Lewin and M. van der Klis, Cambridge Univ. Press.


\bibitem[Gavriil, Kaspi \& Woods(2002)]{gav02} Gavriil, F. P., Kaspi, V. M., \& Woods, P. M., 2002, \nat, 419, 142

\bibitem[Gonzales \& Safi-Harb(2003)]{gonza03} Gonzalez, M., \& Safi-Harb, S., 2003, \apj,  591, L143

\bibitem[Gonzalez et al.(2004)]{gonza04} Gonzalez, M. E., Kaspi, V. M., Lyne, A. G., \& Pivovaroff, M. J., 2004,
\apj, 610, L37

\bibitem[Heger, Langer \& Woosley(2000)]{heger} Heger, A., Langer, N., \& Woosley, S. E. 2000, \apj, 528, 368

\bibitem[Illarionov \& Sunyaev(1975)]{IS75} Illarionov, A. F., \& Sunyaev, R.A.,
1975, \aap , 39, 185.

\bibitem[Li(2003)]{li03} Li, X.-D., 2003, \apj, 596, L199.

\bibitem[Li(2002)]{li02} Li, X.-L, 2002, \apj, 567, 463

\bibitem[Lin, Woosley \& Bodenheimer(1991)]{LWB} Lin, D. N. C., Woosley, S. E., \& Bodenheimer, P. H., 1991, \nat, 353, 827

\bibitem[McLaughlin et al.(2003a)]{mclaugh03a} McLaughlin, M. A., et al.,\ 2003a, in
Young Neutron Stars and Their Environments, IAU Symposium, Vol. 218, 2004,
F. Camilo and B. M. Gaensler, eds. astro-ph/0310455

\bibitem[McLaughlin et al.(2003b)]{mclaugh03b} McLaughlin, M. A., et al.,\ 2003b,
\apj, L135

\bibitem[Marsden et al.(2001a)]{marsden01a} Marsden, D., Lingenfelter, R. E.,
Rothschild R. E., \& Higdon, J. C., 2001a, \apj, 550, 397.

\bibitem[Marsden et al.(2001b)]{marsden01b} Marsden, D., Lingenfelter, R. E.,
\& Rothschild R. E., 2001b, \apj, 547, L45.

\bibitem[Menou et al.(2001)]{menou01} Menou, K., Perna R., \& Hernquist, L. 2001, \apj, 554, L63.

\bibitem[Meyer-Hofmeister(1992)]{meyer92} Meyer-Hofmeister, E., 1992, \aap, 253, 459

\bibitem[Meyer \& Meyer-Hofmeister(1989)]{meyer89} Meyer, F., \& Meyer-Hofmeister, E., 1989,
in \emph{Theory of Accretion Disks}, NATO ARW, eds. F. Meyer et al., Kluwer Academic Publishers, p. 307.

\bibitem[Michel \& Dessler(1981)]{MD81} Michel, F. C., \& Dessler, A. J., 1981, \apj, 251, 654

\bibitem[Mineshige, Nomoto \& Shigeyama(1993)]{mineshige93} Mineshige, S., Nomoto, K., \& Shigeyama, T., 1993, \aap, 267, 95

\bibitem[Mineshige et al.(1997)]{minesh97} Mineshige, S., Nomura, H., Hirose, M., Nomoto, K., \& Suzuki, T., 1997, \apj, 489, 227.

\bibitem[Mineshige, Rees \& Fabian(1991)]{min91} Mineshige, S., Rees, M. J., \& Fabian, A. C., 1991, \mnras, 251, 555

\bibitem[\"{O}gelman \& Alpar(2004)]{OA04} \"{O}gelman, H., \& Alpar, M. A., 2004, \apj, 603, L33

\bibitem[Park(2004)]{park04} Park, S., 2004, Adv. Sp. Res., 33, 386

\bibitem[Pavlov, Sanwal \& Teter(2004)]{pavlov04} Pavlov, G. G., Sanwal, D., \&
Teter, M. A., 2003, in Young Neutron Stars and Their Environments, IAU
Symposium, Vol. 218, 2004, F. Camilo and B. M. Gaensler, eds. astro-ph/0311526

\bibitem[Perna, Hernquist \& Narayan(2000)]{PHN} Perna, R., Hernquist, L., \& Narayan, R., 2000, \apj, 541, 344

\bibitem[Priedhorsky(1986)]{pri86} Priedhorsky, W., 1986, \apj, 306, L97

\bibitem[Qiao et al.(2003)]{qiao03} Qiao, G. J., Xue, Y. Q., Xu, R. X., Wang, H. G., \& Xiao, B. W., 2003, \aap, 407, L25

\bibitem[Romanova et al.(2004)]{romanova04} Romanova, M. M., Ustyugova, G. V., Koldoba, A. V., \& Lovelace, R. V. E. 2004 \apj 616, L151

\bibitem[Shaham \& Tavani(1986)]{sha86} Shaham, J., \& Tavani, M., 1986, IAU Symposium 125, The
Origin and Evolution of Neutron Stars, ed. D. J. Helfand and J. H. Huang (Dordrecht: Reidel), p. 199

\bibitem[Shi \& Xu(2003)]{shi03} Shi, Y., \& Xu, R.X., 2003, \apj, 596, L75

\bibitem[Shvartsman(1970a)]{shv70a} Shvartsman, V. F., 1970, Radiofizika, 13, 1852

\bibitem[Shvartsman(1970b)]{shv70b} Shvartsman, V. F., 1970, Astron. Zh., 47, 660

\bibitem[Spruit \& Taam(1993)]{ST93} Spruit, H. C, \& Taam, R. E., 1993, \apj, 402, 593

\bibitem[Tam et al.(2004)]{tam04} Tam, C. R., Kaspi, V. M., van Kerkwijk, M. H., \& Durant, M., \apj, 617, L53

\bibitem[Treves \& Bocci(1987)]{tre87} Treves, A. \& Bocci, F., 1987, \mnras, 225, Short Communication, 39

\bibitem[Vivekanand \& Narayan(1981)]{vive81} Vivekanand, M., \& Narayan, R., 1981, J. Astrophys. Astr. 2, 315

\bibitem[Vranesevic et al.(2004)]{vra04} Vranesevic, N. et al. 2004, \apj, 617, L139

\bibitem[Woods \& Thompson(2004)]{WT04} Woods, \& P. M., Thompson, C., 2004, in
"Compact Stellar X-ray Sources", eds. W.H.G. Lewin and M. van der Klis, Cambridge Univ. Press.

\bibitem[Xu, Wang \& Qiao(2003)]{xuwang} Xu, R. X, Wang, H. G., Qiao, G. J., 2003, Chin. Phys. Lett., 2, 314

\end{thebibliography}
\end{document}